# SAT Solving for Argument Filterings[⋆]

Michael Codish[1], Peter Schneider–Kamp[2], Vitaly Lagoon[3],
René Thiemann[2], and Jürgen Giesl[2]

[1] Dept. of Computer Science, Ben-Gurion University, Israel, mcodish@cs.bgu.ac.il
[2] LuFG Informatik 2, RWTH Aachen, Germany,
{psk,thiemann,giesl}@informatik.rwth-aachen.de
[3] Dept. of Computer Science and Software Engineering, University of Melbourne,
Australia, lagoon@cs.mu.oz.au

**Abstract.** This paper introduces a propositional encoding for lexicographic path orders in connection with dependency pairs. This facilitates the application of SAT solvers for termination analysis of term rewrite systems based on the dependency pair method. We address two main inter-related issues and encode them as satisfiability problems of propositional formulas that can be efficiently handled by SAT solving: (1) the combined search for a lexicographic path order together with an *argument filtering* to orient a set of inequalities; and (2) how the choice of the argument filtering influences the set of inequalities that have to be oriented. We have implemented our contributions in the termination prover AProVE. Extensive experiments show that by our encoding and the application of SAT solvers one obtains speedups in orders of magnitude as well as increased termination proving power.

## 1 Introduction

In recent work [5], Codish *et al.* introduce a propositional encoding of lexicographic path orders (LPO) and demonstrate that SAT solving can drastically speedup the solving of LPO termination problems. The key idea is that the encoding of a term rewrite system (TRS) $\mathcal{R}$ is satisfiable if and only if $\mathcal{R}$ is LPO-terminating and that each model of the encoding indicates a particular LPO which orients the rules in $\mathcal{R}$. However, lexicographic path orders on their own are too weak for many interesting termination problems and hence LPO is typically combined with more sophisticated termination proving techniques. One of the most popular and powerful such techniques is the *dependency pair* (DP) method [2]. Essentially, for any TRS the DP method generates a set of inequalities between terms. If one can find a well-founded order satisfying these inequalities, then termination is proved. A main advantage of the DP method is that it permits the use of orders which need not be monotonic. This allows the application of lexicographic path orders combined with *argument filterings*.

For every function symbol $f$, an argument filtering $\pi$ specifies which parts of a term $f(\ldots)$ may be eliminated before comparing terms. In recent refinements

---

[⋆] Supported by the Deutsche Forschungsgemeinschaft DFG under grant GI 274/5-1.

of the DP method [9,20], the choice of $\pi$ also influences the set of *usable rules* which contribute to the inequalities that have to be oriented. As stated in [14], "the dependency pairs method derives much of its power from the ability to use argument filterings to simplify constraints". However, argument filterings represent a severe bottleneck for the automation of dependency pairs, as the search space for argument filterings is enormous.

This paper extends the approach of [5] by providing a propositional encoding which combines the search for an LPO with the search for an argument filtering. This extension is non-trivial as the choice of an argument filtering $\pi$ influences the structure of the terms in the rules as well as the set of rules which contribute to the inequalities that need to be oriented. The key idea is to combine all of the constraints on $\pi$ which influence the definition of the LPO and the definition of the usable rules and to encode these constraints in SAT. This encoding captures the synergy between precedences on function symbols and argument filterings. In our approach there exist an argument filtering $\pi$ and an LPO which orient a set of inequalities if and only if the encoding of the inequalities is satisfiable. Moreover, each model of the encoding corresponds to a suitable argument filtering and a suitable LPO which orient the inequalities.

After presenting in Sect. 2 the necessary preliminaries on LPO and on the DP method, Sect. 3 extends the approach of [5] to consider argument filterings. Sect. 4 shows how to extend this encoding to take into account the influence of an argument filtering on the set of usable rules. In Sect. 5 we describe the implementation of our results in the termination prover AProVE [12] and provide extensive experimental evidence which indicates speedups in orders of magnitude. We conclude in Sect. 6.

## 2 Preliminaries

This section briefly describes the starting points for the rest of the paper: propositional encodings for lexicographic path orders [5,17] and the dependency pair framework [2,10,14]. We refer to [3] for further details on term rewriting.

We assume an algebra of terms constructed over given sets of symbols $\mathcal{F}$ and variables $\mathcal{V}$. Let $>_{\mathcal{F}}$ denote a (strict or non-strict) partial order on $\mathcal{F}$ (a so-called *precedence*) and let $\approx_{\mathcal{F}}$ denote the corresponding equivalence relation. We denote by $\sim$ the equality of terms up to equivalence of symbols. Observe that if $>_{\mathcal{F}}$ is strict then $\approx_{\mathcal{F}}$ and $\sim$ are the identity of symbols and terms respectively. Each precedence $>_{\mathcal{F}}$ on the symbols induces a lexicographic path order on terms.

**Definition 1 (LPO [16]).** *The lexicographic path order $\succ_{LPO}$ on terms induced by the partial order $>_{\mathcal{F}}$ is defined as $s = f(s_1, \ldots, s_n) \succ_{LPO} t$ if and only if one of the following holds:*

1. *$t = g(t_1, \ldots, t_m)$ and $s \succ_{LPO} t_j$ for all $1 \leq j \leq m$ and either*
   *(i) $f >_{\mathcal{F}} g$    or    (ii) $f \approx_{\mathcal{F}} g$ and $\langle s_1, \ldots, s_n \rangle \succ_{LPO}^{lex} \langle t_1, \ldots, t_m \rangle$; or*
2. *$s_i \succsim_{LPO} t$ for some $1 \leq i \leq n$.*

Here $\succ_{LPO}^{lex}$ is the lexicographic extension of $\succ_{LPO}$ to tuples of terms and $\succsim_{LPO}$ is the union of $\succ_{LPO}$ and $\sim$.

The classical approach to prove termination of a TRS $\mathcal{R}$ is to find a *reduction order* $\succ$ which orients ($\ell \succ_{LPO} r$) all of the rules $\ell \to r$ in $\mathcal{R}$. A reduction order is an order which is well-founded, monotonic, and stable (closed under contexts and substitutions). In practice, most reduction orders amenable to automation are *simplification orders* [7], i.e., they contain the embedding relation $\succ_{emb}$.

The lexicographic path order is one of the most prominent simplification orders and raises the associated decision problem: For terms $s$ and $t$, does there exist a precedence $>_\mathcal{F}$ such that $s \succ_{LPO} t$ holds? This decision problem comes in two flavours: "strict-LPO" and "quasi-LPO" depending on whether $>_\mathcal{F}$ is required to be strict or not. In [13], the authors observe that finding $>_\mathcal{F}$ such that $s \succ_{LPO} t$ is tantamount to solving a constraint obtained by unfolding the definition of $s \succ_{LPO} t$.

As an example, let $\mathcal{F} = \{-, +, *\}$. Then there exists a strict precedence such that $-(x+y) \succ_{LPO} (-x) * (-y)$ if and only if the *partial order constraint* $(- >_\mathcal{F} *) \vee ((+ >_\mathcal{F} *) \wedge (+ >_\mathcal{F} -))$ has a solution. In [17] the authors show how such constraints can be encoded into propositional formulas. These formulas are satisfiable if and only if there exists a suitable partial order. A substantially improved encoding from such partial order constraints into propositional formulas is presented in [5].

However, it is well known that lexicographic path orders on their own are not very powerful for proving termination.

*Example 2.* Consider the following TRS $\mathcal{R}$ for division on natural numbers [2].

$$\mathsf{minus}(x, 0) \to x \quad (1) \qquad \mathsf{quot}(0, \mathsf{s}(y)) \to 0 \quad (3)$$
$$\mathsf{minus}(\mathsf{s}(x), \mathsf{s}(y)) \to \mathsf{minus}(x, y) \quad (2) \qquad \mathsf{quot}(\mathsf{s}(x), \mathsf{s}(y)) \to \mathsf{s}(\mathsf{quot}(\mathsf{minus}(x, y), \mathsf{s}(y))) \quad (4)$$

Rules (1) - (3) can easily be oriented using an LPO, but rule (4) cannot. To see this, observe that if we instantiate $y$ by $\mathsf{s}(x)$, we obtain $\mathsf{quot}(\mathsf{s}(x), \mathsf{s}(\mathsf{s}(x))) \prec_{emb} \mathsf{s}(\mathsf{quot}(\mathsf{minus}(x, \mathsf{s}(x)), \mathsf{s}(\mathsf{s}(x))))$. Thus, no simplification order can show termination of $\mathcal{R}$. This drawback was the reason for developing more powerful approaches like the dependency pair method.

The dependency pair framework [10] is a modular reformulation and improvement of Arts and Giesl's dependency pair approach [2] which was also inspired by related work in [4,14]. To ease readability, the following presentation is slightly simplified yet sufficient to state the contributions of this paper. For further details on the dependency pair framework see [10].

For a term rewrite system $\mathcal{R}$ over the symbols $\mathcal{F}$, the set of *defined* symbols $\mathcal{D}_\mathcal{R} \subseteq \mathcal{F}$ is the set of all root symbols of left-hand sides of $\mathcal{R}$. With each defined symbol $f \in \mathcal{D}_\mathcal{R}$ we extend the signature $\mathcal{F}$ by a fresh *tuple symbol* $F$. For each rule $f(s_1, \ldots, s_n) \to r$ in a term rewrite system $\mathcal{R}$ and for each subterm $g(t_1, \ldots, t_m)$ of $r$ with $g \in \mathcal{D}_\mathcal{R}$, $F(s_1, \ldots, s_n) \to G(t_1, \ldots, t_m)$ is a dependency pair, intuitively indicating that a function call to $f$ may lead to a function call to $g$. The set of dependency pairs of $\mathcal{R}$ is denoted $DP(\mathcal{R})$.

*Example 3.* Recall the term rewrite system from Ex. 2. The defined symbols are
minus and div and there are three dependency pairs:

$$\mathsf{MINUS}(\mathsf{s}(x), \mathsf{s}(y)) \to \mathsf{MINUS}(x, y) \tag{5}$$

$$\mathsf{QUOT}(\mathsf{s}(x), \mathsf{s}(y)) \to \mathsf{MINUS}(x, y) \tag{6}$$

$$\mathsf{QUOT}(\mathsf{s}(x), \mathsf{s}(y)) \to \mathsf{QUOT}(\mathsf{minus}(x, y), \mathsf{s}(y)) \tag{7}$$

The main result underlying the dependency pair method states that a term rewrite system $\mathcal{R}$ is terminating if and only if there is no infinite (minimal) $\mathcal{R}$-*chain* of its dependency pairs $DP(\mathcal{R})$ [2]. In other words, there is no infinite sequence of dependency pairs $s_1 \to t_1, s_2 \to t_2, \ldots$ from $DP(\mathcal{R})$ such that for all $i$ there is a substitution $\sigma_i$ where $t_i \sigma_i$ is terminating with respect to $\mathcal{R}$ and $t_i \sigma_i \to_{\mathcal{R}}^* s_{i+1} \sigma_{i+1}$. To prove absence of such infinite chains automatically, we consider so-called *dependency pair problems*. A dependency pair problem $(\mathcal{P}, \mathcal{R})$ is a pair of term rewrite systems $\mathcal{P}$ and $\mathcal{R}$ and poses the question: "Is there an infinite $\mathcal{R}$-chain of dependency pairs from $\mathcal{P}$?" The goal is to solve the dependency pair problem $(DP(\mathcal{R}), \mathcal{R})$ in order to determine termination of $\mathcal{R}$.

Termination techniques now operate on dependency pair problems and are called *DP processors*. Formally, a DP processor *Proc* takes a dependency pair problem as input and returns a new dependency pair problem which then has to be solved instead. A processor *Proc* is *sound* if for all dependency pair problems $(\mathcal{P}, \mathcal{R})$ where $Proc(\mathcal{P}, \mathcal{R}) = (\mathcal{P}', \mathcal{R})$, there is an infinite $\mathcal{R}$-chain of pairs from $\mathcal{P}'$ whenever there is an infinite $\mathcal{R}$-chain of pairs from $\mathcal{P}$. Soundness of a DP processor is required to prove termination and in particular, to conclude that there is no infinite $\mathcal{R}$-chain if $Proc(\mathcal{P}, \mathcal{R}) = (\varnothing, \mathcal{R})$.

So termination proofs in the DP framework start with the initial DP problem $(DP(\mathcal{R}), \mathcal{R})$. Then the DP problem is simplified repeatedly by sound DP processors. If one reaches the DP problem $(\varnothing, \mathcal{R})$, then termination is proved. In the following, we present one of the most important processors of the framework, the so-called *reduction pair processor*. Additional processors are described in [10].

For a DP problem $(\mathcal{P}, \mathcal{R})$, the reduction pair processor generates inequality constraints which should be satisfied by a *reduction pair* $(\succsim, \succ)$ [18] where $\succsim$ is reflexive, transitive, monotonic, and stable and $\succ$ is a stable well-founded order compatible with $\succsim$ (i.e., $\succsim \circ \succ \subseteq \succ$ or $\succ \circ \succsim \subseteq \succ$). However, $\succ$ need not be monotonic. A typical choice for a reduction pair $(\succsim, \succ)$ is to use simplification orders in combination with *argument filterings* [2] (we adopt notation of [18]).

**Definition 4 (Argument Filtering).** *An* argument filtering $\pi$ *maps every n-ary function symbol to an argument position $i \in \{1, \ldots, n\}$ or to a (possibly empty) list $[i_1, \ldots, i_p]$ with $1 \leq i_1 < \cdots < i_p \leq n$. An argument filtering $\pi$ induces a mapping from terms to terms:*

$$\pi(t) = \begin{cases} t & \text{if } t \text{ is a variable} \\ \pi(t_i) & \text{if } t = f(t_1, \ldots, t_n) \text{ and } \pi(f) = i \\ f(\pi(t_{i_1}), \ldots, \pi(t_{i_p})) & \text{if } t = f(t_1, \ldots, t_n) \text{ and } \pi(f) = [i_1, \ldots, i_p] \end{cases}$$

For a relation $\succ$ on terms, let $\succ^\pi$ be the relation where $s \succ^\pi t$ holds if and only if $\pi(s) \succ \pi(t)$. An argument filtering with $\pi(f) = i$ is called *collapsing* on $f$.

Arts and Giesl show in [2] that if $(\succsim, \succ)$ is a reduction pair and $\pi$ is an argument filtering then $(\succsim^\pi, \succ^\pi)$ is also a reduction pair. In particular, we focus on reduction pairs of the form $(\succsim^\pi_{LPO}, \succ^\pi_{LPO})$ to prove termination of examples like Ex. 2 where the direct application of simplification orders fails.

The constraints generated by the reduction pair processor require that (a) all dependency pairs in $\mathcal{P}$ are weakly or strictly decreasing and, (b) all *usable* rules $\mathcal{U}(\mathcal{P}, \mathcal{R})$ are weakly decreasing. Here, a rule $f(\ldots) \to r$ from $\mathcal{R}$ is *usable* if $f$ occurs in the right-hand side of a dependency pair from $\mathcal{P}$ or of a usable rule. In Ex. 2, the symbols occurring in the right-hand sides of the dependency pairs $(5) - (7)$ are MINUS, QUOT, s, and minus. Therefore the minus-rules $(1)$ and $(2)$ are usable. Since the right-hand sides of the minus-rules do not contain additional symbols, these are in fact all of the usable rules. Hence, the quot-rules $(3)$ and $(4)$ are not usable.

As shown in [15,20], under certain conditions on the reduction pair, Restriction (b) ensures that in chains $s_1 \to t_1, s_2 \to t_2, \ldots$ with $t_i \sigma_i \to^*_{\mathcal{R}} s_{i+1} \sigma_{i+1}$, we have $t_i \sigma_i \succsim s_{i+1} \sigma_{i+1}$. The required conditions hold in particular for any reduction pair constructed using simplification orders and argument filterings and specifically for $(\succsim^\pi_{LPO}, \succ^\pi_{LPO})$. Hence, the strictly decreasing pairs of $\mathcal{P}$ cannot occur infinitely often in chains. This enables the processor to delete such pairs from $\mathcal{P}$.

In the following, for any term rewrite system $\mathcal{Q}$ and relation $\succ$, we denote $\mathcal{Q}_\succ = \{s \to t \in \mathcal{Q} \mid s \succ t\}$.

**Theorem 5 (Reduction Pair Processor).** *Let $(\succsim, \succ)$ be a reduction pair for a simplification order $\succ$ and let $\pi$ be an argument filtering. Then the following DP processor Proc is sound.*

$$Proc(\mathcal{P}, \mathcal{R}) = \begin{cases} (\mathcal{P} \setminus \mathcal{P}_{\succ^\pi}, \mathcal{R}) & \text{if } \mathcal{P}_{\succ^\pi} \cup \mathcal{P}_{\succsim^\pi} = \mathcal{P} \text{ and } \mathcal{R}_{\succsim^\pi} \supseteq \mathcal{U}(\mathcal{P}, \mathcal{R}) \\ (\mathcal{P}, \mathcal{R}) & \text{otherwise} \end{cases}$$

*Example 6.* For the term rewrite system of Ex. 2, according to Thm. 5 we search for a reduction pair solving the following inequality constraints.

$$\mathsf{minus}(x, 0) \succsim x \tag{8}$$
$$\mathsf{minus}(\mathsf{s}(x), \mathsf{s}(y)) \succsim \mathsf{minus}(x, y) \tag{9}$$
$$\mathsf{MINUS}(\mathsf{s}(x), \mathsf{s}(y)) \succ_{(\succsim)} \mathsf{MINUS}(x, y) \tag{10}$$
$$\mathsf{QUOT}(\mathsf{s}(x), \mathsf{s}(y)) \succ_{(\succsim)} \mathsf{MINUS}(x, y) \tag{11}$$
$$\mathsf{QUOT}(\mathsf{s}(x), \mathsf{s}(y)) \succ_{(\succsim)} \mathsf{QUOT}(\mathsf{minus}(x, y), \mathsf{s}(y)) \tag{12}$$

Following Thm. 5, all of the inequalities (10)-(12) which are strictly decreasing can be removed. To solve the inequalities we may take $(\succsim^\pi_{LPO}, \succ^\pi_{LPO})$ where $\pi(\mathsf{minus})=1$, $\pi(\mathsf{s})=\pi(\mathsf{MINUS})=\pi(\mathsf{QUOT})=[1]$, and where $\succsim_{LPO}$ and $\succ_{LPO}$ are induced by the partial order $\mathsf{QUOT} >_\mathcal{F} \mathsf{MINUS}$. For this choice, inequalities (10)-(12) are all strict and hence removed by the reduction pair processor. This results in the new DP problem $(\varnothing, \mathcal{R})$ which proves termination of Ex. 2.

We conclude this brief description of the dependency pair framework with a statement of the central *decision problem* associated with argument filterings, LPO, and dependency pairs:

> For a given dependency pair problem $(\mathcal{P}, \mathcal{R})$, does there exist a reduction pair $(\succsim_{LPO}^{\pi}, \succ_{LPO}^{\pi})$ for some argument filtering $\pi$ and lexicographic path order induced by some partial order $>_\mathcal{F}$ such that all rules in $\mathcal{P}$ and in $\mathcal{R}$ are weakly decreasing and at least one rule in $\mathcal{P}$ is strictly decreasing?

In the following section we show how to encode constraints like "$s \succ_{LPO}^{\pi} t$" and "$s \succsim_{LPO}^{\pi} t$" as propositional formulas. Given such an encoding enables to encode the decision problem stated above as a SAT problem. Based on the solution of the SAT problem one can then identify the dependency pairs which can be removed from $\mathcal{P}$.

## 3 Encoding LPO and Argument Filtering

In this section we consider lexicographic path orders with argument filterings and the corresponding decision problem. Consider first a naive brute force approach. For any given argument filtering $\pi$ we generate the formula

$$\bigwedge_{s \to t \in \mathcal{U}(\mathcal{P},\mathcal{R})} \pi(s) \succsim_{LPO} \pi(t) \ \wedge\ \bigwedge_{s \to t \in \mathcal{P}} \pi(s) \succsim_{LPO} \pi(t) \ \wedge\ \bigvee_{s \to t \in \mathcal{P}} \pi(s) \succ_{LPO} \pi(t) \quad (13)$$

The constraints "$\pi(s) \succsim_{LPO} \pi(t)$" and "$\pi(s) \succ_{LPO} \pi(t)$" can be encoded as described in Sect. 2. Then SAT solving can search for an LPO satisfying (13) for the given filtering $\pi$. However, this approach is hopelessly inefficient, potentially calling the SAT solver for each of the exponentially many argument filterings.

A contribution of this paper is to show instead how to encode the argument filterings into the propositional formula and delegate the search for an argument filtering to the SAT solver. In this way, the SAT solver is only called once with an encoding of Formula (13) and it can search for an argument filtering and for a precedence at the same time. This is clearly advantageous, since the filtering and the precedence highly influence each other.

So our goal is to encode constraints like "$s \succ_{LPO}^{\pi} t$" (or "$s \succsim_{LPO}^{\pi} t$") into propositional formulas such that every model of the encoding corresponds to a concrete filtering $\pi$ and precedence $>_\mathcal{F}$ which satisfy "$s \succ_{LPO}^{\pi} t$" (or "$s \succsim_{LPO}^{\pi} t$"). We first provide an explicit definition which then provides the basis for specifying partial order and argument filtering constraints, satisfaction of which give "$s \succ_{LPO}^{\pi} t$" (or "$s \succsim_{LPO}^{\pi} t$"). The essential differences with Definition 1 are two: each of the two cases of Definition 1 are refined to consider the effect of $\pi$; and we use the weak version $\succsim_{LPO}^{\pi}$ of the order instead of equivalence on terms.

**Definition 7 (LPO modulo $\pi$).** *Let $>_\mathcal{F}$ be a (strict or non-strict) precedence and let $\pi$ be an argument filtering on $\mathcal{F}$. Let $x$ denote a variable.*

(I) The induced lexicographic path order $\succ_{LPO}^{\pi}$ on terms is defined as follows: $s = f(s_1, \ldots, s_n) \succ_{LPO}^{\pi} t$ if and only if one of the following holds:
1. $t = g(t_1, \ldots, t_m)$ and
   (a) $\pi(g) = j$ and $s \succ_{LPO}^{\pi} t_j$; or
   (b) $\pi(f) = [i_1, \ldots, i_p]$, $\pi(g) = [j_1, \ldots, j_q]$, $s \succ_{LPO}^{\pi} t_j$ for all $j \in [j_1, \ldots, j_q]$, and either (i) $f >_{\mathcal{F}} g$, or
   (ii) $f \approx_{\mathcal{F}} g$ and $\langle s_{i_1}, \ldots, s_{i_p}\rangle \succ_{LPO}^{\pi,lex} \langle t_{j_1}, \ldots, t_{j_q}\rangle$; or
2. (a) $\pi(f) = i$ and $s_i \succ_{LPO}^{\pi} t$; or
   (b) $\pi(f) = [i_1, \ldots, i_p]$ and for some $i \in [i_1, \ldots, i_p]$, $(s_i \succsim_{LPO}^{\pi} t)$.

(II) For tuples of terms we define $\langle s_1, \ldots, s_n\rangle \succ_{LPO}^{\pi,lex} \langle t_1, \ldots, t_m\rangle$ iff $n > 0$ and
   (a) $m = 0$ or
   (b) $m > 0$ and $((s_1 \succ_{LPO}^{\pi} t_1) \vee ((s_1 \succsim_{LPO}^{\pi} t_1) \wedge \langle s_2, \ldots, s_n\rangle \succ_{LPO}^{\pi,lex} \langle t_2, \ldots, t_m\rangle))$.

(III) $\succsim_{LPO}^{\pi}$ and $\succsim_{LPO}^{\pi,lex}$ are defined in an analogous way to $\succ_{LPO}^{\pi}$ and $\succ_{LPO}^{\pi,lex}$:
   (a) replacing $\succ_{LPO}^{\pi}$ by $\succsim_{LPO}^{\pi}$ in (I) 1(a) and 2(a); and
   (b) adding the case $x \succsim_{LPO}^{\pi} g(t_1, \ldots, t_m)$ iff $\pi(g) = j$ and $x \succsim_{LPO}^{\pi} t_j$ and the case $x \succsim_{LPO}^{\pi} x$ to (I); and
   (c) replacing $\succ_{LPO}^{\pi,lex}$ by $\succsim_{LPO}^{\pi,lex}$ in (I),(II) and adding $\langle\rangle \succsim_{LPO}^{\pi,lex} \langle\rangle$ to (II).

It follows directly from Definitions 1, 4, and 7 that for all terms $s$ and $t$ we have $s \succ_{LPO}^{\pi} t \Leftrightarrow \pi(s) \succ_{LPO} \pi(t)$ and $s \succsim_{LPO}^{\pi} t \Leftrightarrow \pi(s) \succsim_{LPO} \pi(t)$.

The decision problem associated with Def. 7 is stated as follows: For terms $s$ and $t$, does there exist a partial order $>_{\mathcal{F}}$ and an argument filtering $\pi$ such that $s \succ_{LPO}^{\pi} t$ resp. $s \succsim_{LPO}^{\pi} t$ holds. This problem again comes in two flavours: "strict-LPO" and "quasi-LPO" depending on if $>_{\mathcal{F}}$ is required to be strict or not. Our aim is to encode these decision problems as constraints on $>_{\mathcal{F}}$ and $\pi$, similar to the encoding of $s \succ_{LPO} t$ as a partial order constraint in Sect. 2. The difference is that now we have two types of constraints: constraints on the partial order $>_{\mathcal{F}}$ and constraints on the argument filtering $\pi$. To express constraints on argument filterings we use atoms of the following forms: "$\pi(f) = i$" to constrain $\pi$ to map $f$ to the value $i$; "$\pi(f) \supseteq i$" to constrain $\pi$ to map $f$ either to a list containing $i$ or to $i$ itself; and "$list(\pi(f))$" to constrain $\pi$ to map $f$ to a list. So "$list(\pi(f))$" means that $\pi$ is not collapsing on $f$.

Each of the cases (I) - (III) in Def. 7 induces an encoding to constraints on partial orders and argument filterings. In the following definition, we illustrate the encoding of $s \succ_{LPO}^{\pi} t$ for the case of strict-LPO with argument filterings. The encoding for $s \succsim_{LPO}^{\pi} t$ and the encodings for quasi-LPO are defined in a similar way. In the following definition, $\tau_{1a}$, $\tau_{1b}$ and $\tau_2$ are the parts of the encoding corresponding to cases 1(a), 1(b) and 2(a-b) in Def. 7 (I).

**Definition 8 (Encoding strict-LPO with Argument Filterings).** *The strict-LPO encoding of $s \succ_{LPO}^{\pi} t$ is a mapping $\tau$ from pairs of terms $s$ and $t$ to constraints defined by the rules depicted in Fig. 1 (where $x$ denotes a variable).*

*Example 9.* Consider the first arguments of QUOT in dependency pair (7). Using Def. 8, after simplification of conjunctions, disjunctions, and implications with

*Encoding I:*

$$\tau(s \succ_{LPO}^{\pi} t) = \tau_{1a}(s \succ_{LPO}^{\pi} t) \bigvee \tau_{1b}(s \succ_{LPO}^{\pi} t) \bigvee \tau_2(s \succ_{LPO}^{\pi} t)$$

$$\tau_{1a}(x \succ_{LPO}^{\pi} t) = \tau_{1b}(x \succ_{LPO}^{\pi} t) = \tau_2(x \succ_{LPO}^{\pi} t) = \tau_{1a}(s \succ_{LPO}^{\pi} x) = \tau_{1b}(s \succ_{LPO}^{\pi} x) = false$$

$$\tau_{1a}(s \succ_{LPO}^{\pi} g(t_1, \ldots, t_m)) = \bigvee_{1 \leq j \leq m} \Big((\pi(g) = j) \bigwedge \tau(s \succ_{LPO}^{\pi} t_j)\Big) \text{ for non-variable } s$$

$$\tau_{1b}(f(s_1, \ldots, s_n) \succ_{LPO}^{\pi} g(t_1, \ldots, t_m)) = list(\pi(f)) \bigwedge list(\pi(g)) \bigwedge (f >_{\mathcal{F}} g) \bigwedge$$
$$\bigwedge_{1 \leq j \leq m} (\pi(g) \sqsupseteq j) \to \tau(f(s_1, \ldots, s_n) \succ_{LPO}^{\pi} t_j) \text{ for } f \neq g$$

$$\tau_{1b}(f(s_1, \ldots, s_n) \succ_{LPO}^{\pi} f(t_1, \ldots, t_n)) = list(\pi(f)) \bigwedge$$
$$\tau(\langle s_1, \ldots, s_n \rangle \succ_{LPO,f}^{\pi,lex} \langle t_1, \ldots, t_n \rangle) \bigwedge$$
$$\bigwedge_{1 \leq j \leq n} (\pi(f) \sqsupseteq j) \to \tau(f(s_1, \ldots, s_n) \succ_{LPO}^{\pi} t_j)$$

$$\tau_2(f(s_1, \ldots, s_n) \succ_{LPO}^{\pi} t) = \bigvee_{1 \leq i \leq n} \Big((\pi(f) = i) \bigwedge \tau(s_i \succ_{LPO}^{\pi} t)\Big) \quad \vee$$
$$\left( list(\pi(f)) \wedge \bigvee_{1 \leq i \leq n} (\pi(f) \sqsupseteq i) \wedge \tau(s_i \succsim_{LPO}^{\pi} t) \right)$$

*Encoding II:*

$$\tau(\langle s_i, \ldots, s_n \rangle \succ_{LPO,f}^{\pi,lex} \langle t_i, \ldots, t_n \rangle) = \textsf{false if } n = 0 \text{ else}$$
$$((\pi(f) \sqsupseteq i) \bigwedge \tau(s_i \succ_{LPO}^{\pi} t_i)) \bigvee$$
$$\Big( ( (\pi(f) \sqsupseteq i) \to \tau(s_i \succsim_{LPO}^{\pi} t_i) ) \bigwedge \tau(\langle s_{i+1}, \ldots, s_n \rangle \succ_{LPO,f}^{\pi,lex} \langle t_{i+1}, \ldots, t_n \rangle) \Big)$$

**Fig. 1.** Encoding LPO with Argument Filterings

*true* and *false* we obtain:

$$\tau(\mathsf{s}(x) \succ_{LPO}^{\pi} \mathsf{minus}(x,y)) = (\pi(\mathsf{minus}) = 1 \wedge list(\pi(\mathsf{s})) \wedge \pi(\mathsf{s}) \sqsupseteq 1)$$
$$\vee (list(\pi(\mathsf{s})) \wedge list(\pi(\mathsf{minus})) \wedge (\mathsf{s} >_{\mathcal{F}} \mathsf{minus}) \wedge$$
$$(\pi(\mathsf{minus}) \sqsupseteq 1 \to list(\pi(\mathsf{s})) \wedge \pi(\mathsf{s}) \sqsupseteq 1) \wedge \neg(\pi(\mathsf{minus}) \sqsupseteq 2))$$

Thus, $\mathsf{s}(x) \succ_{LPO}^{\pi} \mathsf{minus}(x,y)$ holds if and only if minus is collapsed to its first argument and s is not filtered or if s and minus are not collapsed, s is greater than minus in the precedence, and the second argument of minus is filtered and whenever minus keeps the first argument then s keeps the first argument, too.

We proceed to describe how partial order and argument filtering constraints are transformed into propositional logic. The propositional encoding of partial order constraints is presented in more detail in [5].

Let $|\mathcal{F}| = m$. Then the basic idea is to interpret the symbols in $\mathcal{F}$ as indices in a partial order taking finite domain values from the set $\{1, \ldots, m\}$. Each symbol $f \in \mathcal{F}$ is thus modeled as $\langle f_k, \ldots, f_1 \rangle$ with $f_k$ the most significant bit and $k = \lceil \log_2 m \rceil$. The binary value of $\langle f_k, \ldots, f_1 \rangle$ represents the position of $f$ in the partial order. One may of course have $\langle f_k, \ldots, f_1 \rangle = \langle g_k, \ldots, g_1 \rangle$ for $f \neq g$, if, for example, a (strict) partial order imposes no order between $f$ and $g$, or if a non-strict partial order imposes $f \approx_\mathcal{F} g$. Constraints of the form $(f >_\mathcal{F} g)$ or $(f \approx_\mathcal{F} g)$ on $\mathcal{F}$ are interpreted as constraints on indices and it is straightforward to encode them in $k$-bit arithmetic: A constraint of the form $(f \approx_\mathcal{F} g)$ is encoded in $k$ bits by

$$\|(f \approx_\mathcal{F} g)\|_k = \bigwedge_{1 \leq i \leq k} (f_i \leftrightarrow g_i).$$

A constraint of the form $(f >_\mathcal{F} g)$ is encoded in $k$-bits by

$$\|(f >_\mathcal{F} g)\|_k = \begin{cases} (f_1 \wedge \neg g_1) & \text{if } k = 1 \\ (f_k \wedge \neg g_k) \vee ((f_k \leftrightarrow g_k) \wedge \|(f > g)\|_{k-1}) & \text{if } k > 1 \end{cases}$$

To encode argument filtering constraints, we associate with each symbol $f \in \mathcal{F}$ of arity $n$ the propositional variables $list_f$ (which is true if and only if $\pi$ is not collapsing on $f$) and $arg_f^1, \ldots, arg_f^n$ (which indicate which arguments of $f$ remain after filtering by $\pi$). We impose for each $f \in \mathcal{F}$ of arity $n$ a constraint of the form $\neg list_f \rightarrow \bigoplus_{1 \leq i \leq n} arg_f^i$ where $\bigoplus_{1 \leq i \leq n} arg_f^i$ specifies that exactly one of the variables $arg_f^i$ is $true$ and the rest are $false$. The argument filtering constraints are then encoded as follows: $\|list(\pi(f))\| = list_f$; $\|\pi(f) \supseteq i\| = arg_f^i$; and $\|\pi(f) = i\| = \neg list_f \wedge arg_f^i$.

*Example 10.* Consider the encoding in Ex. 9 which still contains partial order constraints and argument filtering constraints. Using the above encoding for these constraints, we obtain the following propositional formula. Since there are only $m = 2$ symbols s and minus, we choose $k = 1$ and encode the partial order constraint $(s >_\mathcal{F} \text{minus})$ as $(s_1 \wedge \neg \text{minus}_1)$.

$$\begin{aligned}
\|\tau(\mathsf{s}(x) \succ_{LPO}^\pi \mathsf{minus}(x,y))\| &= (\neg list_\mathsf{minus} \wedge arg_\mathsf{minus}^1 \wedge list_\mathsf{s} \wedge arg_\mathsf{s}^1) \\
&\vee (list_\mathsf{s} \wedge list_\mathsf{minus} \wedge (\mathsf{s}_1 \wedge \neg \mathsf{minus}_1) \wedge \\
&\quad (arg_\mathsf{minus}^1 \rightarrow list_\mathsf{s} \wedge arg_\mathsf{s}^1) \wedge \neg arg_\mathsf{minus}^2)
\end{aligned}$$

## 4 Argument Filterings and Usable Rules

Recent improvements of the DP method [9,20] significantly reduce the number of rules required to be weakly decreasing in the reduction pair processor of Thm. 5. We first recapitulate the improved reduction pair processor and then adapt our propositional encoding accordingly.

The idea is that one can restrict the set of usable rules by taking the argument filtering into account: in right-hand sides of dependency pairs or rules, an occurrence of $f$ in the $i$-th argument of $g$ will never be the cause to introduce a usable $f$-rule if the argument filtering eliminates $g$'s $i$-th argument. For instance, when taking $\pi(\mathsf{QUOT}) = [2]$ in Ex. 2, the right-hand side of the *filtered* dependency pairs do not contain minus anymore. Thus, no rule is considered usable. In Def. 11, we define these restricted usable rules for a term $t$ (initially corresponding to the right-hand side of a dependency pair). Here, we make the TRS $\mathcal{R}$ explicit to facilitate a straightforward encoding in Def. 14 afterwards.

**Definition 11 (Usable Rules modulo $\pi$ [9,20]).** *Let $\mathcal{R}$ be a TRS and $\pi$ an argument filtering. For any function symbol $f$, let $Rls_\mathcal{R}(f) = \{\ell \to r \in \mathcal{R} \mid \text{root}(\ell) = f\}$. For any term $t$, the* usable rules $\mathcal{U}_\pi(t, \mathcal{R})$ *modulo $\pi$ are given by:*

$$\mathcal{U}_\pi(x, \mathcal{R}) = \varnothing \quad \text{for all variables } x$$
$$\mathcal{U}_\pi(f(t_1, \ldots, t_n), \mathcal{R}) = Rls_\mathcal{R}(f) \cup$$
$$\bigcup_{\ell \to r \in Rls_\mathcal{R}(f)} \mathcal{U}_\pi(r, \mathcal{R} \setminus Rls_\mathcal{R}(f)) \cup$$
$$\bigcup_{\pi(f) \ni i} \mathcal{U}_\pi(t_i, \mathcal{R} \setminus Rls_\mathcal{R}(f))$$

*For a set of dependency pairs $\mathcal{P}$, let $\mathcal{U}_\pi(\mathcal{P}, \mathcal{R}) = \bigcup_{s \to t \in \mathcal{P}} \mathcal{U}_\pi(t, \mathcal{R})$.*

We now refine the reduction pair processor of Thm. 5 to consider usable rules modulo $\pi$.

**Theorem 12 (Reduction Pair Processor modulo $\pi$ [20]).** *Let $(\succsim, \succ)$ be a reduction pair for a simplification order $\succ$ and let $\pi$ be an argument filtering. Then the following DP processor Proc is sound.*

$$Proc(\mathcal{P}, \mathcal{R}) = \begin{cases} (\mathcal{P} \setminus \mathcal{P}_{\succ^\pi}, \mathcal{R}) & \text{if } \mathcal{P}_{\succ^\pi} \cup \mathcal{P}_{\succsim^\pi} = \mathcal{P} \text{ and } \mathcal{R}_{\succsim^\pi} \supseteq \mathcal{U}_\pi(\mathcal{P}, \mathcal{R}) \\ (\mathcal{P}, \mathcal{R}) & \text{otherwise} \end{cases}$$

*Example 13.* Consider the following TRS (together with the minus-rules (1), (2))

| | | | |
|---|---|---|---|
| $\text{ge}(x, 0) \to \text{true}$ | (14) | $\text{div}(x, y) \to \text{if}(\text{ge}(x, y), x, y)$ | (17) |
| $\text{ge}(0, \text{s}(y)) \to \text{false}$ | (15) | $\text{if}(\text{true}, \text{s}(x), \text{s}(y)) \to \text{s}(\text{div}(\text{minus}(x, y), \text{s}(y)))$ | (18) |
| $\text{ge}(\text{s}(x), \text{s}(y)) \to \text{ge}(x, y)$ | (16) | $\text{if}(\text{false}, x, \text{s}(y)) \to 0$ | (19) |

The usable rules are the minus- and ge-rules since minus occurs in the right-hand side of the dependency pair $\text{IF}(\text{true}, \text{s}(x), \text{s}(y)) \to \text{DIV}(\text{minus}(x, y), \text{s}(y))$ resulting from rule (18) and ge occurs in the dependency pair $\text{DIV}(x, y) \to \text{IF}(\text{ge}(x, y), x, y)$ resulting from rule (17). However, if one chooses the argument filtering $\pi(\text{DIV}) = \pi(\text{GE}) = \pi(\text{MINUS}) = \pi(\text{s}) = [1]$, $\pi(\text{IF}) = [2]$, and $\pi(\text{minus}) = 1$, then the ge-rules are no longer usable since ge does not occur in the right-hand side of the filtered dependency pair $\text{DIV}(x) \to \text{IF}(x)$. Now Thm. 12 only requires the filtered minus-rules and the dependency pairs to be decreasing.

As demonstrated in [9,20] and confirmed by the experiments described in Sect. 5, introducing argument filterings to the specification of usable rules results in a significant gain of termination proving power. However, Thm. 12 is not straightforward to automate using SAT solvers. The technique of Sect. 3 assumes a given set of inequalities which is then encoded to a propositional formula. The problem with Thm. 12 is that that the set of inequalities to be oriented depends on the chosen argument filtering. Hence, the search for an argument filtering should be combined with the computation of the usable rules. Once again, the alternative brute force enumeration of argument filterings is hopelessly inefficient. Therefore, we modify the encoding of the inequalities in Formula (13) to consider for every rule $\ell \to r \in \mathcal{R}$, the condition under which $\ell \to r$ is usable.

Only under this condition one has to require the inequality $\pi(\ell) \succsim \pi(r)$. To this end, instead of encoding formula (13) we encode the following formula.

$$\underbrace{\bigwedge_{\ell \to r \in \mathcal{U}_\pi(\mathcal{P},\mathcal{R})} \ell \succsim^\pi_{LPO} r}_{(a)} \wedge \underbrace{\bigwedge_{s \to t \in \mathcal{P}} s \succsim^\pi_{LPO} t}_{(b)} \wedge \underbrace{\bigvee_{s \to t \in \mathcal{P}} s \succ^\pi_{LPO} t}_{(c)} \qquad (13')$$

The subformulas $(b)$ and $(c)$ are identical to those in Formula (13) and are encoded as a conjunction and disjunction of encodings of the forms $\tau(s \succsim^\pi_{LPO} t)$ and $\tau(s \succ^\pi_{LPO} t)$ using Def. 8. The definition of the usable rules in Def. 11 now induces the following encoding of subformula $(a)$ as a propositional formula $\omega(\mathcal{P},\mathcal{R})$.[1] Here, we reuse the encoding for "$\pi(f) \supseteq i$" as presented in Sect. 3. Moreover we introduce a new propositional variable $u_f$ for every defined function symbol $f$ of $\mathcal{U}(\mathcal{P},\mathcal{R})$ which indicates whether $f$'s rules are usable.

**Definition 14 (Encoding Usable Rules modulo Argument Filtering).**
*For a term $t$ and a TRS $\mathcal{R}$ the formula $\omega(t, \mathcal{R})$ is defined as follows:*

$$\begin{aligned}
\omega(x, \mathcal{R}) &= \text{true} && \text{for } x \in \mathcal{V} \\
\omega(f(t_1, \ldots, t_n), \mathcal{R}) &= \bigwedge_{1 \le i \le n} (\pi(f) \supseteq i \to \omega(t_i, \mathcal{R})) && \text{for } f \notin \mathcal{D}_\mathcal{R} \\
\omega(f(t_1, \ldots, t_n), \mathcal{R}) &= u_f \wedge && \text{for } f \in \mathcal{D}_\mathcal{R} \\
&\quad \bigwedge_{\ell \to r \in Rls_\mathcal{R}(f)} \omega(r, \mathcal{R} \setminus Rls_\mathcal{R}(f)) \wedge \\
&\quad \bigwedge_{1 \le i \le n} (\pi(f) \supseteq i \to \omega(t_i, \mathcal{R} \setminus Rls_\mathcal{R}(f)))
\end{aligned}$$

*For a set of dependency pairs $\mathcal{P}$, let*

$$\omega(\mathcal{P}, \mathcal{R}) = \left(\bigwedge_{s \to t \in \mathcal{P}} \omega(t, \mathcal{R})\right) \wedge \left(\bigwedge_{f \in \mathcal{D}_{\mathcal{U}(\mathcal{P},\mathcal{R})}} u_f \to \left(\bigwedge_{\ell \to r \in Rls_\mathcal{R}(f)} \tau(\ell \succsim^\pi_{LPO} r)\right)\right).$$

For a DP problem $(\mathcal{P}, \mathcal{R})$ we encode the formula (13'). Every model of this encoding corresponds to a precedence $>_\mathcal{F}$ and an argument filtering $\pi$ satisfying the constraints of the improved reduction pair processor from Thm. 12. Thus, we can now use SAT solving to automate Thm. 12 as well.

*Example 15.* Consider the TRS $\mathcal{R}$ from Ex. 13. Using the encoding of Def. 14, for $\mathcal{P} = DP(\mathcal{R})$ we obtain:

$$\begin{aligned}
\omega(\mathcal{P}, \mathcal{R}) = &(\pi(\mathsf{DIV}) \supseteq 1 \to u_\mathsf{minus}) \wedge (\pi(\mathsf{IF}) \supseteq 1 \to u_\mathsf{ge}) \wedge \\
&(u_\mathsf{minus} \to (\tau(\mathsf{minus}(x, 0) \succsim^\pi_{LPO} x) \wedge \tau(\mathsf{minus}(\mathsf{s}(x), \mathsf{s}(y)) \succsim^\pi_{LPO} \mathsf{minus}(x, y)))) \wedge \\
&\quad (u_\mathsf{ge} \to (\tau(\mathsf{ge}(x, 0) \succsim^\pi_{LPO} \mathsf{true}) \wedge \tau(\mathsf{ge}(0, \mathsf{s}(y)) \succsim^\pi_{LPO} \mathsf{false}) \wedge \\
&\qquad \tau(\mathsf{ge}(\mathsf{s}(x), \mathsf{s}(y)) \succsim^\pi_{LPO} \mathsf{ge}(x, y))))
\end{aligned}$$

---

[1] The definition of $\omega$ can easily be adapted to more advanced definitions of usable rules as well, cf. e.g. [2,9,11].

## 5 Implementation and Experiments

The propositional encodings for LPO with argument filterings and for the reduction pair processors described in Sect. 3 and 4 have been fully implemented and integrated within the termination prover AProVE available from [6]. This Java implementation consists of the following main components: **(a)** An encoder from DP problems to formulas with partial order and argument filtering constraints (ca. 1700 lines). **(b)** A propositional encoder for partial order constraints following [5] and for argument filtering constraints (ca. 300 lines). **(c)** Interfaces to several SAT solvers (ca. 300 lines). For the scope of this paper all results are obtained using the MiniSAT solver [8]. For the translation to conjunctive normal form (CNF) we used the implementation of Tseitin's algorithm [21] offered by SAT4J [19] - a freely available Java implementation of MiniSAT.

Our implementation uses several optimizations to minimize encoding size:

1. We apply basic simplification axioms for true and false as well as standard Boolean simplifications to flatten nested conjunctions and disjunctions.
2. When building the formulas top-down, at each point we maintain the sets of atomic constraints (partial order and argument filtering) that must be true and false from this point on. This information is then applied to simplify all constraints generated below (in the top-down process) and to prune the encoding process.
3. We memo and identify identical subformulas in the propositional encodings and instead of representing formulas as trees we represent them as directed acyclic graphs (or Boolean circuits). This decreases the size of the representation considerably. For instance, consider the constraint from Ex. 9. Already in this tiny example, the subformula $list(\pi(\mathsf{s})) \wedge \pi(\mathsf{s}) \supseteq 1$ occurs twice, since it results from the encoding of both $\mathsf{s}(x) \succ_{LPO}^{\pi} x$ and $\mathsf{s}(x) \succ_{LPO}^{\pi} y$.

Optimization (2) typically reduces the number of propositional variables in the resulting CNF by a factor of at least 2. Optimizations (1) and (3) together further reduce the number of propositional variables by a typical factor of 10.

To evaluate our new SAT-based implementation, we performed extensive experiments to compare it with the corresponding methods in the current non-SAT-based implementations of AProVE [12] and of the Tyrolean Termination Tool (TTT) [15]. As shown in the annual *International Competition of Termination Tools* 2004 and 2005, AProVE and TTT are the two most powerful tools for termination analysis of term rewriting. For our experiments, both AProVE and TTT are configured to consider all argument filterings.

We ran the three tools on all 773 TRSs from the *Termination Problem Data Base*. This is the collection of examples from the annual competition of termination tools [1]. For the experiments, the TTT analyzer is applied via its web interface and runs on a Xeon 2.24GHz dual-CPU platform. The AProVE analyzer and our new SAT-based analyzer are run on an AMD Athlon 64 at 2.2 GHz.

Apart from the reduction pair processor, we also used the *dependency graph processor* [2,10,14], which is the other main processor of the dependency pair framework. This processor is used to split up dependency pair problems into smaller ones. As AProVE and TTT use slightly different techniques for estimating dependency graphs in the dependency graph processor and they run on different machines, their performance is not directly comparable.

Tables 1 and 2 summarize the results using the DP processors based on Thm. 5 and Thm. 12 respectively. The tools are indicated as: TTT,[2] APR (AProVE) and SAT (AProVE with our SAT-based encoding). For each of the experiments we consider reduction pairs based on *strict-* and *quasi-*LPO. Each of the experiments was performed with a time-out of 60 seconds (corresponding to the way tools are evaluated in the annual competition) and with a time-out of 10 minutes. We indicate by "*Yes*", "*Fail*", and "*RL*" the number of TRSs for which proving termination with the given technique succeeds, fails, or encounters a resource limit (time-out or exhausts memory). Finally, we give the total time in seconds for analyzing all 773 examples. Individual runtimes and proof details are available from our empirical evaluation web site [6].

|      | LPO - 60sec t/o | | | | LPO - 10min t/o | | | | QLPO - 60sec t/o | | | | QLPO - 10min t/o | | | |
|------|---|---|---|---|---|---|---|---|---|---|---|---|---|---|---|---|
| Tool | Yes | Fail | RL | Time | Yes | Fail | RL | Time | Yes | Fail | RL | Time | Yes | Fail | RL | Time |
| TTT  | 268 | 448 | 57 | 4202 | 269 | 465 | 39 | 28030 | 297 | 395 | 81 | 6241 | 297 | 408 | 68 | 43540 |
| APR  | 310 | 358 | 105 | 6936 | 310 | 365 | 98 | 60402 | 320 | 331 | 122 | 7913 | 326 | 341 | 106 | 67764 |
| SAT  | 327 | 446 | 0 | 119 | 327 | 446 | 0 | 119 | 359 | 411 | 3 | 423 | 359 | 414 | 0 | 563 |

**Table 1.** Strict-LPO (left) and Quasi-LPO (right) with the DP processor of Thm. 5

The comparison of the corresponding SAT-based and non-SAT-based configurations in Table 1 shows that the analyzers based on SAT solving with our proposed encoding are faster by orders of magnitude. Moreover, the power (i.e., the number of examples where termination can be proved) also increases substantially in the SAT based configurations. It is also interesting to note that there are almost no time-outs in the SAT-based configurations, whereas the non-SAT-based configurations have many time-outs.[3]

|      | LPO - 60sec t/o | | | | LPO - 10min t/o | | | | QLPO - 60sec t/o | | | | QLPO - 10min t/o | | | |
|------|---|---|---|---|---|---|---|---|---|---|---|---|---|---|---|---|
| Tool | Yes | Fail | RL | Time | Yes | Fail | RL | Time | Yes | Fail | RL | Time | Yes | Fail | RL | Time |
| APR  | 338 | 368 | 67 | 4777 | 341 | 383 | 49 | 33329 | 357 | 323 | 93 | 6100 | 359 | 336 | 78 | 49934 |
| SAT  | 348 | 425 | 0 | 115 | 348 | 425 | 0 | 115 | 380 | 390 | 3 | 435 | 380 | 393 | 0 | 587 |

**Table 2.** Strict-LPO (left) and Quasi-LPO (right) with the DP processor of Thm. 12

Table 2 provides results using the improved reduction pair processor of Thm. 12. Again, the SAT-based configuration is much faster than the corresponding non-SAT-based one. The comparison with Table 1 shows that replac-ing the processor of Thm. 5 by the one of Thm. 12 increases power significantly and has no negative influence on runtimes.

---

[2] TTT offers two algorithms to search for argument filterings. We used the "divide-and-conquer"-algorithm, since it is usually the more efficient one.

[3] To evaluate the optimizations on p. 12, we also tested the SAT-based configuration with strict-LPO and the 10-minute time-out in a version where optimizations (2) and (3) are switched off. Here, the total runtime increases from 119 to 1968 seconds. Thus, optimizations (2) and (3) already decrease the total runtime by a factor of 16.

In both tables, the comparison between strict- and quasi-LPO (of corresponding configurations) shows that quasi-LPO is more powerful but also slower than strict-LPO. However, for the SAT-based analyses, the overall runtimes are still extremely fast in comparison to the non-SAT-based configurations.

Table 3 highlights 5 examples which could not be solved by any tool in the International Termination Competition 2005, whereas the SAT-based configuration proves termination for all 5 in a total of 5.3 seconds. In fact, except for the second example, neither TTT nor AProVE are able to prove termination in their fully automatic mode within 10 minutes. This demonstrates that our encoding advances the state of the art of automated termination analysis. The columns labeled TTT, APR, and SAT indicate for the three tools the analysis times in seconds (to find a proof of termination) and "t/o" indicates a 10 minute timeout. For each of the examples and tools, the time indicated is for the fastest configuration from those described in Tables 1 and 2. For the second and third example, TTT's "divide-and-conquer"-algorithm times out, but its "enumeration"-algorithm (which is usually less efficient) finds a solution within 10 minutes. Therefore, here the runtimes are given in brackets. The last four columns give details on the termination proof with SAT. Column 4 and 5 indicate the number of clauses and the number of literals for the largest CNF which occurred during the proof (ranging over all dependency pair problems encountered). Columns 6 and 7 indicate the time (in milliseconds) for encoding and for SAT solving (the rest of the time is spent for reading and parsing, producing outputs, computing dependency graphs, etc).

| Example | TTT | APR | SAT | # clauses | # literals | encod. time | SAT time |
|---|---|---|---|---|---|---|---|
| Ex26_Luc03b_Z | t/o | t/o | 0.95 | 11335 | 29794 | 117 | 44 |
| Ex2_Luc02a_C | (476.8) | t/o | 1.31 | 7968 | 21385 | 159 | 16 |
| Ex49_GM04_C | ( 25.8) | 44.4 | 1.14 | 6641 | 17654 | 213 | 12 |
| ExSec11_1_Luc02a_C | t/o | t/o | 1.32 | 9187 | 24536 | 159 | 20 |
| ExSec11_1_Luc02a_GM | t/o | t/o | 0.58 | 18158 | 48072 | 221 | 52 |

**Table 3.** Five hard examples: SAT solving increases termination proving power

## 6 Conclusion

In [5] the authors demonstrate the power of propositional encoding and application of SAT solving to LPO termination analysis. This paper extends the SAT-based approach to consider the more realistic setting of dependency pair problems with LPO and argument filtering. The main challenge derives from the strong dependencies between the notions of LPO, argument filterings, and the set of rules which need to be oriented. The key to a solution is to introduce and encode in SAT all of the constraints originating from these notions into a single search process. We introduce such an encoding and through implementation and experimentation prove that it meets the challenge yielding speedups in orders of magnitude over existing termination tools as well as increasing termination proving power. To experiment with our SAT-based implementation and for further details on our experiments please visit our web page at http://aprove.informatik.rwth-aachen.de/eval/SATLPO [6].